\documentclass[twoside]{article}
\usepackage[accepted]{aistats2023}
\usepackage{physics}
\usepackage{url}
\usepackage{fancyhdr}
\usepackage{latexsym,revsymb}
\usepackage{listings}
\usepackage{verbatim}
\usepackage{amsthm}
\usepackage{amsmath,amssymb}
\usepackage{mathtools}
\usepackage{bm}
\usepackage{graphicx}
\usepackage{ascmac}
\usepackage{wrapfig}
\usepackage{enumitem}
\usepackage{tikz}

\usepackage[round]{natbib}

\usetikzlibrary{positioning,fit,calc, arrows.meta,shapes,decorations,automata,backgrounds,petri,topaths,calc,mindmap,trees,positioning,chains,arrows}
\tikzstyle{block} = [draw=black, thick, text width=7cm, minimum height=2.6cm, align=center, rounded corners]  
\tikzstyle{bblock} = [draw=black, thick, text width=7cm, minimum height=3.4cm, align=center, rounded corners]  
\tikzstyle{arrow} = [thick,->,>=stealth]

\newcommand*{\myDots}{\ifmmode\mathellipsis\else.\kern-0.13em.\kern-0.13em.\fi} % touching at \kern-0.1725em

\newcommand{\argmin}{\mathop{\rm arg~min}\limits}

\newcommand{\ten}{\mathsf{T}}

\newcommand{\av}[1]{\left \langle {#1} \right \rangle}
\newcommand{\sgn}[1]{{\rm sgn}\qty(#1)}
\newcommand{\E}{\mathbb{E}}

\DeclareMathOperator*{\Extr}{{\rm Extr}}
\newcommand{\lassox}{\hat{\bm{x}}_\lambda}

\newcommand{\iu}{\mathrm{i}\mkern1mu}
\newcommand{\TP}{{\mathrm{TP}}}
\newcommand{\FP}{{\mathrm{FP}}}

\newcommand{\erfc}{{\rm erfc}}
\newcommand{\Prob}{{\Probability}}
\newcommand{\lfit}[2]{ \bm{\gamma}_{{#1}}( {#2} ) }
\newcommand{\tN}{\tilde{N}}
\newcommand{\tba}{\tilde{\bm{a}}}

\newtheorem{claim}{Claim}
\newtheorem{assumption}{Assumption}

\newtheorem{lemma}{Lemma}

\begin{document}

\twocolumn[

\aistatstitle{Average case analysis of Lasso under ultra-sparse conditions}

\aistatsauthor{ Koki Okajima \And Xiangming Meng \And Takashi Takahashi \And Yoshiyuki Kabashima }

\aistatsaddress{ Department of Physics, The University of Tokyo } ]

\begin{abstract}
  We analyze the performance of the least absolute shrinkage and selection operator (Lasso)
  for the linear model
when the number of regressors $N$ grows larger keeping the true support size $d$ finite, i.e., the ultra-sparse case. 
The result is based on a novel treatment of the non-rigorous replica method 
in statistical physics, which has been applied only to problem settings where 
$N$, $d$ and the number of observations $M$ tend to infinity at the same rate. 
{\color{black}
Our analysis makes it possible to assess the average performance 
of Lasso with Gaussian sensing matrices 
 without assumptions on the scaling of $N$ and $M$, 
 the noise distribution, and the profile of the true signal. }
Under mild conditions on the noise distribution, the analysis also offers 
a lower bound on the sample complexity necessary for partial and perfect 
support recovery when $M$ diverges as $M = O(\log N)$. 
The obtained bound for perfect support recovery is a generalization of 
that given in previous literature, which only considers the case of Gaussian noise {\color{black} and diverging $d$.} 
Extensive numerical experiments strongly support our analysis.

\end{abstract}

\section{Introduction}
An important objective of high dimensional statistics is to extract information in situations where the signal's dimension $N$ 
is overwhelmingly large compared to the accumulated sample size $M$. 
It is crucial to incorporate prior knowledge on the signal structure to reduce the signal space dimension for reliable estimation. 
A particularly common assumption is \textit{sparsity}, which postulates that the true signal 
has few nonzero entries. 
Exploiting this property allows one to obtain robust and interpretable results specifying the few 
relevant variables explaining the retrieved data \citep{Donoho06}. 

For instance, consider the sparse linear regression problem where measurements $\bm{y} \in \mathbb{R}^M$
of the signal $\bm{x}^0 \in \mathbb{R}^N$ with $d$ non-zero components are given by the linear model 
\begin{equation}
  \bm{y} = \vb{A}\bm{x}^0 + \bm{\xi},
\end{equation}
where $\vb{A}\in \mathbb{R}^{M\times N}$ is the sensing matrix, and $\bm{\xi} \in \mathbb{R}^M$ 
is the noise vector distributed according to $p_{\xi}(\bm{\xi})$. 
The most fundamental yet popular sparse signal estimation method is the least absolute shrinkage and selection operator (Lasso) \citep{Tibshirani96}, 
which offers the estimator by solving the following convex program:
\begin{equation}\label{eq:Lasso_estimator}
  \lassox(\vb{A}, \bm{y}) := \argmin_{\bm{x}} \qty(\frac{1}{2}\norm{\vb{A}\bm{x} - \bm{y}}^2 + M \lambda \norm{\bm{x}}_1) ,
\end{equation}
where $\lambda$ is a regularization parameter. Since its 
introduction, this simple $\ell_1$-regularization scheme has been 
successfully adapted as a backbone 
technique for solving a wide variety of sparse estimation problems. 
A particularly interesting question to ask is if one can make any guarantees on the 
performance of Lasso under general scalings of $(N,M,d)$, its dependence on 
$\lambda$, and statistical properties of the noise and true signal. 

A sheer amount of research has been devoted to assessing the performance of Lasso. 
Traditionally, research based on the irrepresentability condition \citep{Meinshausen06,Zhao06} 
has been popular in establishing guarantees in terms of support recovery of 
the sparse signal \citep{Wainwright09,Dossal12, Meinshausen06,Zhang08,Candes09, Zhao06}. 
A different approach based on 
approximate message-passing (AMP) theory \citep{Donoho09_PNAS}, 
and the heuristical replica method \citep{MPV87} from statistical physics has 
focused on assessing the sharp, asymptotic properties of Lasso in the large $N$ and $M$ limit 
under random sensing matrix designs. 
Despite the previous works, the understanding of the Lasso estimator is still limited. 
Analysis based on the irrepresentability condition often offers only scaling guarantees with respect to 
$(N,M,d)$, or statements with strong assumptions on the regularization parameter. Besides, 
the AMP/replica-based analysis has been only limited to linear sparsity, 
i.e. $d / N  = O(1)$ and $M / N  = O(1)$ as $N\to\infty$, 
which may be somewhat unrealistic compared to real-world situations. 

\subsection{Contributions}
In this work, we complement the drawbacks in both the irrepresentability condition approach and 
AMP / replica approach by
theoretically analyzing the average performance of 
Lasso when $d = O(1)$, i.e. the \textit{ultra-sparse} case \citep{Donoho92,Bhadra17}, 
which is a more typical situation 
in certain applications such as materials informatics \citep{App_Ghiringhelli15,App_Kim16, App_Pilania16}. 
Moreover, our result offers a necessary condition for support recovery in the limit $N, M \to \infty$. 
Specifically, our contributions are summarized as follows:
\begin{itemize}
  \item We provide a new way to apply the replica method in the ultra-sparsity regime. This 
  is done by explicitly handling the correlations and finite-size effects acting on the 
  active set ${\rm supp}(\bm{x}^0) = \qty{i\ |\ x_i^0, \neq 0\ i = 1, \cdots, N }$, which is otherwise ignored in conventional analysis (Section 2.1, Claim \ref{conj:FE}). 
  \item Using this enhanced replica method, we precisely evaluate the average property of Lasso under ultra-sparsity and
   standard Gaussian matrix design, i.e. each element of $\vb{A}$ is i.i.d. according to a standard Gaussian distribution. 
  This provides an extension to previous results derived from the AMP theory and the replica method, 
  where linear sparsity is necessary for the analysis (Section 2.2, Claim \ref{claim:performance_measure}). 
  \item We derive a necessary condition for partial support recovery 
  ${\rm supp} (\hat{\vb{x}}_\lambda(\vb{A}, \vb{y})) \subseteq {\rm supp} (\bm{x}^0)$ 
  under some mild conditions (Assumption \ref{main_assumption}). 
  Specifically, the number of false positives, and subsequently the 
  model misselection probability vanishes only if 
  $M > \alpha_C \log N$ for $N \to \infty$. 
  {\color{black}This constant 
  $\alpha_C$ is determined by the mean prediction error of an oracle
  (Section 2.3, Claim \ref{theorem_main}, \ref{theorem_main2}).}
  \item In addition to partial support recovery, the analysis also provides a necessary condition for perfect support recovery 
  ${\rm supp} (\hat{\vb{x}}_\lambda(\vb{A}, \vb{y})) = {\rm supp} (\bm{x}^0)$,
   which generalizes the sample complexity bound given by \citet{Wainwright09}
    for i.i.d. Gaussian noise distributions in the limit $d \to \infty$ to 
    more general noise distributions under constant $d$ (Section 2.3, Claim \ref{corollary_perfect}). 
  \item We demonstrate that our theory agrees well with experiment by conducting extensive numerical simulations (Section 3).
\end{itemize}
Note that all of the results are derived from the enhanced replica method, 
which is yet to be proven rigorously; hence the statements are presented as claims. 
\subsection{Related Work}

\paragraph*{Irrepresentability Condition.}
As aforementioned, the irrepresentability condition, first introduced by \citet{Meinshausen06} and \citet{Zhao06}, 
has been an important cornerstone, as it establishes a sufficient condition for perfect support 
recovery.  
This condition indicates whether the covariates, i.e. the columns of $\vb{A}$, are linearly independent enough 
to be distinguishable from one another, and hence variable selection is relatively feasible.  
It has been revealed that Lasso is an ``optimal" support estimator in the sublinear regime $d = o(N)$, 
i.e. Lasso has its success/failure threshold for sample complexity in the same order as 
the informational-theoretical one \citep{Fletcher09,Wainwright09IT}.
However, little is known about the constants 
involved in these conditions. \citet{Wainwright09} provided necessary and sufficient 
conditions for 
perfect support recovery under random Gaussian matrices {\color{black}
for diverging $d$.} 
This is a simple and explicit bound which depends on the regularization parameter 
and intensity of the noise, which is restricted to i.i.d. Gaussian. 
{\color{black} Focusing on the case $d = O(1)$,
 \citet{Dossal12} derived sufficient conditions for partial and perfect support recovery 
 under deterministic noise, 
 whose bound is similar to the one given in \citet{Wainwright09}. 
 }

\paragraph*{AMP theory.}
A particular line of work has aimed in assessing the properties of Lasso under 
general random matrix designs via careful analysis of the dynamical behavior of the AMP algorithm \citep{Kabashima03, Donoho09_PNAS,Takahashi20}, 
whose convergence point coincides with \eqref{eq:Lasso_estimator} in the large $N$ limit.
Rather than establishing inequality bounds or conditions, 
the objective is to establish sharp results on the Lasso for a 
random instance of $(\vb{A}, \bm{y})$. 
Although analysis is limited to linear sparsity regime, 
powerful and precise results have been proven rigorously
 under this framework \citep{Bayati12}. 
For instance, \citet{Su17} and \cite{Wang20} determine
 the possible rate of false positives and true positives 
achievable under certain settings, 
which can be obtained by solving a small set of nonlinear equations. 
Nevertheless, the analysis does not give insight on support recovery, since 
this is impossible in the linear sparsity regime \citep{Fletcher09, Wainwright09IT}. 

\paragraph*{Replica method.} 
Results similar to those from AMP theory have also been derived by 
 using the non-rigorous replica method in statistical mechanics. 
Unlike AMP theory, which is based on a convergence analysis of a particular algorithm, the 
replica method aims at directly calculating the average over $(\vb{A}, \bm{y})$ 
of a cumulant generating function for some probability distribution, i.e. of the form 
$K_\phi(t) = \E_{\vb{A}, \bm{y}}\ \log \int \dd \bm{x}\ e^{t\phi(\bm{x})} p(\bm{x} | \vb{A}, \bm{y})$. 
This calculation is often encountered in the field of statistics, where one is interested in the average behavior 
of a statistical model. 
 While lacking a complete proof, this method has been successful 
 in predicting the average performance of machine learning and optimization methods 
 under general random designs in the linear sparsity regime \citep{Vehkapera16, Zdeborova16}. 
In fact, under certain assumptions, the average predictions given by the replica method have been 
proven to be consistent with the asymptotic results obtained from AMP theory 
 and other rigorous methods \citep{Stojnic13, Thrampoulidis18}. 
Similar to AMP theory, however, reliable adaptations of 
this method outside linear sparsity are still open problems. 
Previous research such as \citet{Abbara20}, \citet{Meng21_NeurIPS} and \citet{Meng21} analyzed the 
performance of sparse Ising model selection using a variation of the replica method.
 However, this was accomplished through a series of ansatzes which are 
generally difficult to justify theoretically. 

\subsection{Preliminaries}

Here we summarize the notations used in this paper. 
The expression $\norm{\cdot}$ denotes the $\ell_2$ norm. 
The active set $S$ is defined as the support of the $d$-sparse true signal $\bm{x}^0$, $S := {\rm supp}(\bm{x}^0) = \{ i\ |\ x^0_{i} \neq 0, \ i = 1, \cdots, N  \}$. 
Define $\tN := N-d$, the size of the inactive set. 
The matrix $\vb{A}_S$ denotes the submatrix constructed by concatenating the columns of $\vb{A}$ with indices in $S$. 
The vector $\bm{x}^0_{S}$ denotes the subvector of $\bm{x}^0$ with indices in $S$. 
For simplicity, $\bm{x}^0$ is assumed to be a deterministic, although this can be extended to random signals trivially. 
The expression $\E_{\vb{A}, \bm{y}}$ denotes the average over the 
joint probability with respect to the pair $(\vb{A}, \bm{y})$, i.e. 
\begin{gather*}
  \E_{\vb{A}, \bm{y}}(\cdots) \\
  = \int \dd \bm{y} \dd \vb{A} \dd \bm{\xi}  p_{\xi}(\bm{\xi}) (\cdots) \frac{e^{-\frac{1}{2}{\rm Tr}{\vb{A}^\ten \vb{A}}}}{(2\pi)^{(NM/2)}} \delta(\bm{y} - \vb{A} \bm{x}^0 - \bm{\xi}),
\end{gather*}
where $\delta(\cdot)$ denotes the Dirac delta function. 
The definition of $\E_{\vb{A}_S, \bm{y}}$ follows straightforwardly from the above. 
Also, define $D\bm{z}$ as the standard Gaussian measure, $D\bm{z} = \dd \bm{z} e^{-\norm{\bm{z}}^2/2}/(2\pi)^{n/2} $ for $\bm{z} \in \mathbb{R}^n$. 
Given $(\vb{A}, \bm{x}^0, \bm{y})$ and regularization parameter $\lambda$, 
the oracle Lasso estimator is defined as $\hat{\bm{x}}_{\lambda}(\vb{A}_S, \bm{y})$, which is the Lasso 
estimator with the true support identified beforehand. It is also convenient to define the oracle Lasso fit, defined by 
$\bm{\gamma}_\lambda(\bm{y}) := \vb{A}_S\hat{\bm{x}}_{\lambda}(\vb{A}_S, \bm{y})$, with its 
dependence on $\vb{A}_S$ suppressed for convenience. 
{\color{black}
Given configuration $(\vb{A}, \bm{x}^0, \bm{y})$, and regularization parameter $\lambda$, 
the number of false positives $\rm FP$ 
and the number of true positives $\rm TP$ of the lasso estimator is defined as 
\begin{align}
  {\rm FP}(\vb{A}, \bm{y}) &= \# \qty{ S^C \cap {\rm supp} (\hat{\bm{x}}_\lambda(\vb{A}, \bm{y}) ) },\\
  {\rm TP}(\vb{A}, \bm{y}) &= \# \qty{ S \cap {\rm supp} (\hat{\bm{x}}_\lambda(\vb{A}, \bm{y}) )}, 
\end{align}
where $S^C$ denotes the complement of set $S$ from $\qty{1, \cdots, N}$. 
Without confusion, the dependence on $(\vb{A}_S, \bm{y})$ is suppressed for convenience.  
}

We say that an event $A$ holds with asymptotically high probability (w.a.h.p.) if there exists a constant 
$c > 0$ such that $\Prob[A] > 1 - O(N^{-c})$. %
We also say that $A$ holds with probability approaching one (w.p.a.1) if $\Prob[A] > 1 - o(1)$ as $N\to \infty$. 
\section{Replica analysis}
%\subsection{General Framework}
Define the Boltzmann distribution as 
\begin{equation}
  \label{eq:Boltzmann}
\begin{gathered}
  P_\beta(\bm{x} |\vb{A}, \bm{y} ) \\
  :=Z^{-1}_\beta\qty(\vb{A}, \bm{y}) \exp \qty( -\frac{\beta}{2} \norm{\vb{A}\bm{x} - \bm{y}}^2 - \beta M \lambda \norm{\bm{x}}_1 ),
\end{gathered}
\end{equation}
where $Z_\beta(\vb{A}, \bm{y})$ is the normalization constant. 
Note that in the limit $\beta \to \infty$, \eqref{eq:Boltzmann} converges to a point-wise distribution concentrated 
on the Lasso estimator $\hat{\bm{x}}_\lambda(\vb{A}, \bm{y})$. 
The main objective of our analysis is to calculate the average of the logarithm of $Z_\beta\qty(\vb{A}, \bm{y})$ 
over the random variables $(\vb{A}, \bm{y})$ in the limit $\beta \to \infty$, which is 
called the free energy or the cumulant generating function
\begin{equation}\label{eq:free_energy}
  \mathcal{F} = - \lim_{\beta \to \infty}\beta^{-1} \E_{\vb{A}, \bm{y}} \log Z_\beta(\vb{A}, \bm{y}).
\end{equation}
 The properties of $\hat{\bm{x}}_\lambda(\vb{A}, \bm{y})$ 
averaged over the population of $(\vb{A}, \bm{y})$ can then be assessed by 
taking appropriate derivatives of $\mathcal{F}$. 

Although \eqref{eq:free_energy} is difficult to calculate straightforwardly, 
this can be resolved by using the replica method \citep{MM09, MPV87}, which is based on the following 
equality 
\begin{equation}\label{eq:replica_trick}
  \E_{\vb{A}, \bm{y}} \log Z_\beta(\vb{A}, \bm{y}) = \lim_{n \to +0} \frac{\E_{\vb{A}, \bm{y}}Z^n_\beta(\vb{A}, \bm{y})-1 }{n}.
\end{equation}
Instead of handling the cumbersome $\log$ expression in \eqref{eq:free_energy} directly, 
one calculates the average of the $n$-th power of $Z_\beta$ for $n\in \mathbb{N}$, 
analytically continues this expression to $n\in \mathbb{R}$, and finally takes the limit $n \to +0$. 
Based on this replica "trick", it suffices to calculate 
\begin{equation}
  \label{eq:replicated_PF}
\begin{gathered}
  \E_{\vb{A}, \bm{y}}Z^n_\beta(\vb{A}, \bm{y})  = \E_{\vb{A}, \bm{y}} \int \prod_{a= 1}^n \dd \bm{x}^a \\
\exp \qty( -\frac{\beta}{2} \sum_{a = 1}^n \norm{\vb{A}\bm{x}^a - \bm{y}}^2 - \beta M \lambda \sum_{a = 1}^n \norm{\bm{x}^a}_1 ).
\end{gathered}
\end{equation}
 up to the first order of $n$ to take the $n\to +0$ limit in the right hand side of \eqref{eq:replica_trick}.
\subsection{Outline of the derivation}
Here, we only give a brief outline of the derivation; for details, see Supplementary Materials.
Rewriting $\Delta^a_i := x^a_i \ (i \notin S)$, it is convenient to introduce the auxillary variable 
$h_\mu^a \equiv \sum_{i \notin S}A_{\mu i} \Delta_i^a\ (\mu = 1, \cdots , M)$, which accounts for 
the effect from the variables not in the true support in each replica $a$. 
{\color{black} A crucial observation is that $\qty{A_{\mu i}}_{1\leq \mu \leq M, \\ i\notin S}$ 
is statistically independent from $(\vb{A}_S$, $\bm{y})$, which allows the average 
to be taken individually. 
}
 By taking the average over the Gaussian variables
 $\qty{A_{\mu i}}_{1\leq \mu \leq M, \\ i\notin S}$ first, we find that $h_{\mu}^a$ is Gaussian with zero mean 
 and covariance $\E h_{\mu}^ah_\nu^b = \delta_{\mu \nu}\sum_{i\notin S}\Delta_i^a\Delta_i^b$. 
 By assuming the \textit{replica symmetric} (RS) ansatz \citep{MPV87}
 \begin{equation}\label{eq:RS}
  \sum_{i\notin S}\Delta_i^a\Delta_i^b := \begin{cases}
    Q & a = b \\
    Q - \chi / \beta & \text{otherwise}
  \end{cases},
  %Q - \frac{\chi}{\beta} + \frac{\chi}{\beta} \delta_{ab},
 \end{equation}
 the integral for the replicated vectors $\qty{\bm{\Delta}^a}_{a = 1}^n$ over the whole $\mathbb{R}^{\tN \times n}$ space 
 is restricted to a subspace satisfying the constraints \eqref{eq:RS}. More explicitly, 
 one can rewrite \eqref{eq:replicated_PF} as 
 \begin{equation}\label{eq:IL}
  \E_{\vb{A}_S, \bm{y}} \int \prod_{a = 1}^n \dd \bm{\Delta}^a \int \dd Q \dd \chi e^{-\beta M \lambda \sum_{a = 1}^n \norm{\bm{\Delta}^a}_1 } \mathcal{I} \mathcal{L},
 \end{equation}
 where $\mathcal{I}$ corresponds to the contribution from the RS constraint: i.e. 
 \begin{equation}\label{eq:subshell}
  \mathcal{I}:= \prod_{a = 1}^n \delta \qty( Q - \sum_{i\notin S}(\Delta_i^a)^2 ) \prod_{a < b} \delta \qty( Q - \frac{\chi}{\beta} - \sum_{i\notin S}\Delta_i^a \Delta_i^b ),
 \end{equation}
 and $\mathcal{L}$ is the contribution from the second line of \eqref{eq:replicated_PF}, albeit simplified as a result of replica symmetry:
 \begin{equation}
  \label{eq:L}
  \begin{gathered}
    \mathcal{L}:= \int D\bm{z} \qty(\int  \dd\bm{x}_S e^{-M\beta G(\bm{x}_S; \bm{z}) })^n,\\
    G(\bm{x}_S; \bm{z}) := \frac{\norm{\vb{A}_S \bm{x}_S + \sqrt{Q}\bm{z} - \bm{y}}^2}{2M(1+\chi)} + \lambda \norm{\bm{x}_S}_1.
   \end{gathered}
 \end{equation}
 By using the Fourier representation of the delta function, \eqref{eq:subshell} can be further rewritten as 
 \begin{equation}
  \label{eq:I}
  \begin{gathered}
  \mathcal{I} = \int_{-\iu \infty}^{+\iu \infty}\dd \hat{Q} \dd \hat{\chi} e^{\frac{Mn\beta}{2}\qty( Q\hat{Q} +(n-1) \chi \hat{\chi} - n\beta Q\hat{\chi} )  }  \\
  \times \int D \hat{\bm{z}} e^{ -\frac{M\beta \hat{Q}}{2} \sum_{a = 1}^n\norm{\bm{\Delta}^a}^2 + \beta\sqrt{M\hat{\chi}}\bm{z}^\ten \bm{\Delta}^a + o(\beta) } .
\end{gathered}  
\end{equation}
 Using this expression, the integral with respect to $\{\bm{\Delta}^a\}_{1\leq a \leq n}$ in \eqref{eq:IL} can be calculated analytically.
  Performing the saddle point 
 approximation for large $M$ to the integrals with respect to $(Q, \hat{Q}, \chi, \hat{\chi})$, and finally taking the limit $\beta \to \infty$ 
 after $n\to +0$ in \eqref{eq:replica_trick}
 yields the following expression for $\mathcal{F}$.
\begin{claim}
  \label{conj:FE}
  The free energy is given by 
  \begin{equation}
    \label{eq:FE}
    \begin{gathered}
    \mathcal{F} = \E_{\vb{A}_S, \bm{y}}\ \Extr_{\Theta} \Bigg\{ -\frac{Q\hat{Q} - \chi \hat{\chi}}{2} \\
     \hspace{-9pt}- \frac{\tN}{2\hat{Q}} \qty[ (\Lambda + \hat{\chi}){\rm erfc} \qty( \sqrt{\frac{\Lambda}{2\hat{\chi}}}) - \sqrt{ \frac{2\Lambda \hat{\chi}}{\pi} } e^{-\Lambda / 2\hat{\chi}} ]  \\
\hspace{-9pt}+ \int \hspace{-2pt} D \bm{z} \min_{\bm{x}_S} G(\bm{x}_S; \bm{z}) %\qty[ \frac{\norm{ \vb{A}_S\bm{x} - \sqrt{Q} \bm{z} - \bm{y} }^2}{2(1+\chi)} + M\lambda \norm{\bm{x}}_1 ]
 \Bigg\}. 
    \end{gathered}
\end{equation}
  Here, $\Lambda := (M\lambda)^2$, ${\rm erfc}$ is the complementary error function ${\rm erfc}(x) := 2/\sqrt{\pi} \int_x^\infty \dd t e^{-t^2} $,
   and $\Extr$ refers to the extremum condition with respect to $\Theta := (Q, \hat{Q}, \chi, \hat{\chi})$, which are random variables dependent on $(\vb{A}_S, \bm{y})$. 
\end{claim}
Straightforward calculation shows that the extremum conditions are given by 
\begin{align}
  \label{eq:Q}Q &= \frac{\tN}{\hat{Q}^2} \qty[ (\Lambda + \hat{\chi}){\rm erfc} \qty( \sqrt{\frac{\Lambda}{2\hat{\chi}}}) - \sqrt{ \frac{2\Lambda \hat{\chi}}{\pi} } e^{-\frac{\Lambda}{2\hat{\chi}}} ], \\
  \label{eq:chi}\chi &= \frac{\tN}{\hat{Q}}\erfc \qty( \sqrt{\frac{\Lambda}{2\hat{\chi}}}),\\
  \label{eq:Qhat} \hat{Q} & = \frac{M}{1+\chi} - \frac{1}{1+\chi} \int D \bm{z} \ 
  \nabla \vdot \lfit{(1+\chi)\lambda}{\sqrt{Q} \bm{z} + \bm{y}} \nonumber \\
  &= \frac{M - \int D \bm{z} \norm{ \hat{\bm{x}}_{(1+\chi)\lambda}( \vb{A}_S, \sqrt{Q}\bm{z} + \bm{y}  ) }_0 }{1+\chi}, \\
  \label{eq:chih}\hat{\chi} &= \frac{\int D \bm{z}\norm{ \lfit{(1+\chi)\lambda}{\sqrt{Q}\bm{z} + \bm{y}} - \sqrt{Q}\bm{z} - \bm{y} }^2 }{(1+\chi)^2}, 
\end{align}

where the second equality in \eqref{eq:Qhat} is from Theorem 1 in \citet{Tibshirani12}. 
Note that the dependence of $\Theta$ on $(\vb{A}_S, \bm{y})$ is not explicitly written for sake of simplicity. 
This evaluation of $\mathcal{F}$ reduces the high-dimensional integral over $\vb{A}$ and $\bm{y}$ to an average 
over a four-dimensional extremum problem involving a $M$--dimensional integral with respect to $\bm{z}$, 
which can be numerically computed via iterative substitution and Monte Carlo sampling over $(\bm{A}_S, \bm{y})$ and $\bm{z}$. 

It is interesting to compare our replica analysis in the large $N$ and $M$ limit 
to the ones considering linear sparsity \citep{Kabashima09,Vehkapera16}. 
In linear sparsity, 
the lasso estimator's statistical property can effectively be 
described by a population of $N$ decoupled, independent scalar estimators 
under Gaussian noise with identical intensity as $N \to \infty$. 
This is often referred to as the \textit{decoupling principle} in information theory; see \citet{GuoVerdu05} and \citet{Bayati11} for details. 
In the ultra-sparse case, the elements of the Lasso estimator 
in the active set, consisting of $d = O(1)$ terms, cannot be expected to 
decouple, as finite-size effects of non-Gaussian and correlated nature
 are expected to be significant to describe its profile. 
This is why a $d-$body optimization procedure and 
the average with respect to $(\vb{A}_S, \vb{y})$ appears explicitly in \eqref{eq:FE}. 
On the other hand, the decoupling principle is implicitly employed for the 
$\tN$ non-active variables conditioned on $(\vb{A}_S, \bm{y})$. More explicitly, 
for each configuration of $(\vb{A}_S, \bm{y})$, each element of 
the non-active Lasso estimator is statistically equivalent to 
\begin{equation}
  \label{eq:decoupled}
  \begin{gathered}
  (\hat{\bm{x}}_\lambda(\vb{A}, \bm{y}))_{i\notin S} \sim g_{\lambda}(\hat{Q}, \sqrt{\hat{\chi}}z_i) \\
  = \min_{x} \qty(\frac{\hat{Q}}{2}x^2 - \sqrt{\hat{\chi}} z_ix + M\lambda \abs{x} ),
  \end{gathered}
\end{equation}
where $z_i$ are i.i.d. according to $\mathcal{N}(0,1)$. Note that the decoupling principle, 
rigorously proven under AMP theory, does not necessarily need 
$N$ and $M$ to diverge at the same rate \citep{Rush18}.

\subsection{Performance assessment of Lasso}
  The free energy allows convenient evaluation of averages of certain functions of the estimator. 
  More explicitly, for a function $\Psi:\mathbb{R}^N \to \mathbb{R}$, its average with respect to the 
  Boltzmann distribution \eqref{eq:Boltzmann} and $(\vb{A}, \bm{y})$ is given by 
  \begin{equation}
    \label{eq:final_average}
    \begin{gathered}
      \av{\Psi(\bm{x})}:= \lim_{\beta \to \infty} \E_{\vb{A}, \bm{y}} \int \dd \bm{x} P_\beta(\bm{x} | \vb{A}, \bm{y})\Psi(\bm{x})\\
      = -\lim_{\beta \to \infty} \lim_{h \to 0} \pdv{h} \beta^{-1} \E_{\vb{A}, \bm{y}} \log Z_\beta (\vb{A}, \bm{y}; h\Psi),
   \end{gathered}    
  \end{equation}
  where 
  \begin{equation}
    \begin{gathered}
    Z_\beta(\bm{A}, \vb{y}; h\Psi)  := \int \dd {\bm{x}} e^{-\frac{\beta}{2}\norm{\vb{A}\bm{x} - \bm{y}}^2 - \beta M\lambda \norm{\bm{x}}_1 - \beta h \Psi(\bm{x})}.
    \end{gathered}
  \end{equation}
  For a class of functions $\Psi$, the above can be calculated trivially, 
  which we state in the following claim:
  \begin{claim}[Average with respect to active and inactive sets]\label{claim:performance_measure}
    For arbitrary functions $\psi:\mathbb{R} \to \mathbb{R}$ and $\Psi:\mathbb{R}^d \to \mathbb{R}$, we have 
    \begin{equation}\label{eq:non-active}
      \av{\sum_{i\notin S} \psi( x_i ) } = \tN \E_{\bm{A}_S, \bm{y}} \int D z \psi(g_\lambda( \hat{Q}, \sqrt{\hat{\chi}}z)), 
    \end{equation}
    and
    \begin{equation}\label{eq:active}
      \av{ \Psi(\bm{x}_S) } = \E_{\rm eff} \ \Psi( \hat{\bm{x}}_{(1+\chi)\lambda}( \vb{A}_S,\sqrt{Q}\bm{z} + \bm{y}  ) ),
    \end{equation}
    where $\E_{\rm eff} := \E_{\vb{A}_S, \bm{y}}\int D \bm{z}$, and 
    $(Q, \hat{Q}, \hat{\chi}, \chi)$ is given by the solution of the extremum conditions 
    \eqref{eq:Q}--\eqref{eq:chih} for each $(\vb{A}_S, \bm{y})$. 
    In particular, performance measures such as the average of true positives ($\TP$), false positives ($\FP$) and $\ell_2$ error $\epsilon_x := \norm{\bm{x}_\lambda(\vb{A}, \bm{y})  - \bm{x}^0}^2$
is given by 
\begin{align}
  \av{\TP} &= \E_{\rm eff}\norm{ \hat{\bm{x}}_{(1+\chi)\lambda}(\vb{A}_S,  \sqrt{Q}\bm{z} + \bm{y}) }_0, \label{eq:replica_TP}\\
  \av{\FP} &= \tN \E_{\vb{A}_S, \bm{y}} {\rm erfc} \qty( \sqrt{\frac{\Lambda}{2\hat{\chi}}}), \label{eq:replica_FP}\\
  \av{\epsilon_x} %&:= \E_{\vb{A}, \bm{y}} \norm{\bm{x}_\lambda(\vb{A}, \bm{y})  - \bm{x}^0}^2 \nonumber \\
  &=  \E_{\rm eff} \qty(Q +  \norm{ \hat{\bm{x}}_{(1+\chi)\lambda}(\vb{A}_S,\sqrt{Q}\bm{z} + \bm{y} )- \bm{x}^0_{S} }^2). \label{eq:replica_l2}
\end{align}
\end{claim}
\subsection{Necessary condition for support recovery}
\begin{figure*}[h]
  \begin{centering}
  \includegraphics[width = 17cm]{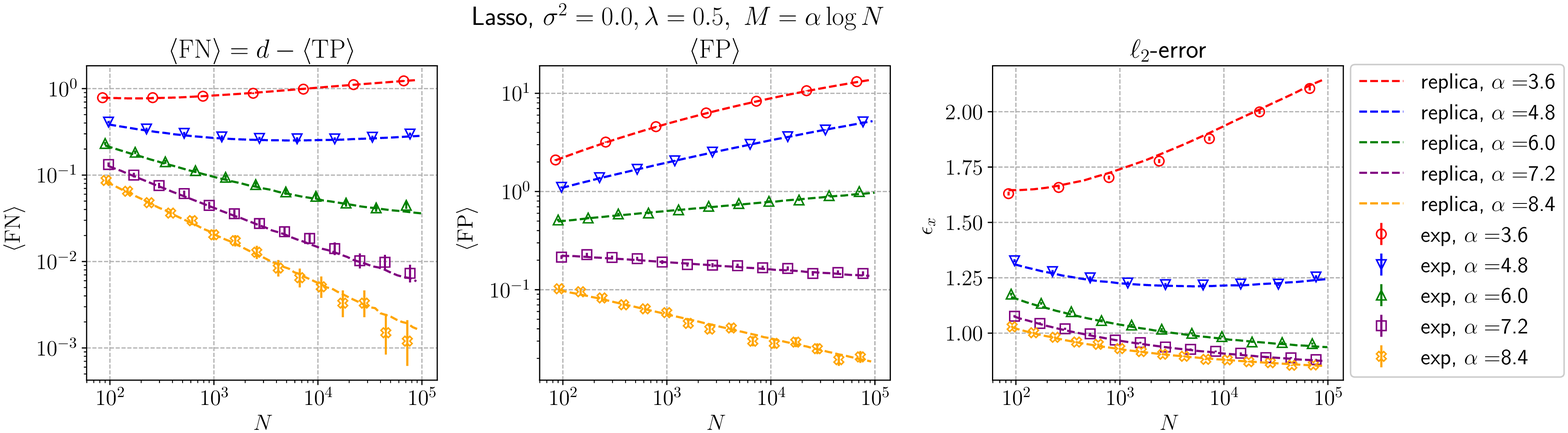} 
\includegraphics[width = 17cm]{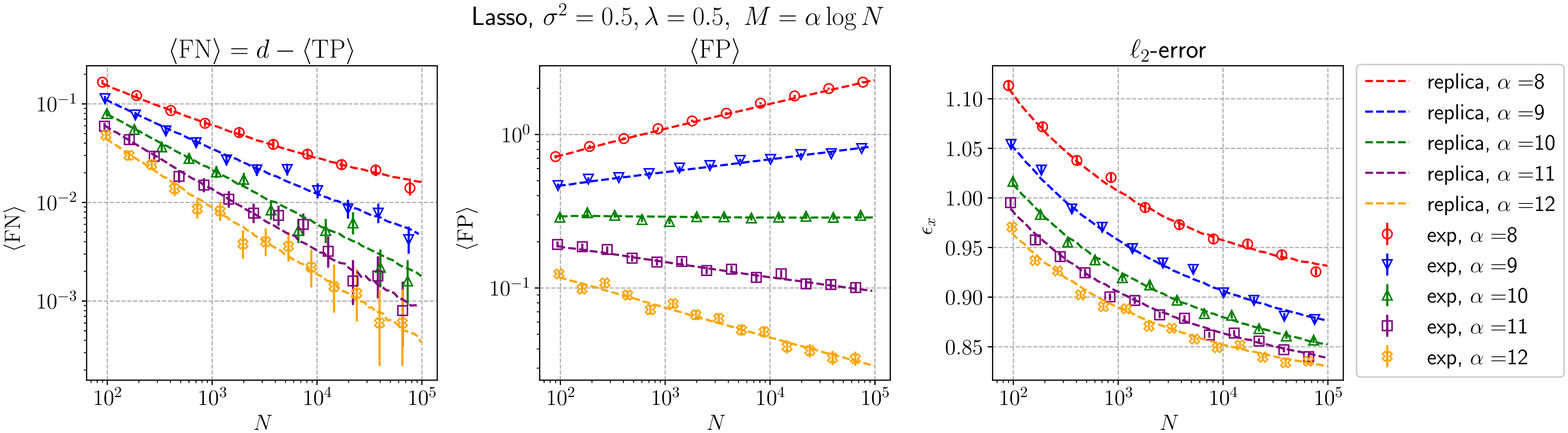}
    \end{centering}
    \caption{Average values of false negatives ($d - \TP$), false positives and $\ell_2$ error for 
    $(\lambda, \sigma^2) = (0.5, 0.0)$ (upper panels) and $(\lambda, \sigma^2) = (0.5, 0.5)$ (lower panels) with $M$ 
    given by $M = \alpha \log N$. 
    Error bars for $\av{\rm FN}$ and $\av{\rm FP}$ represent 
    the 95\% interval of the mean, assuming that
     the samples from the $10^4$ experimental runs follow a binomial distribution. 
     Error bars for $\ell_2$-error represent the standard error obtained from $10^4$ experimental runs. 
    }
  \label{fig:lasso_general}
  \end{figure*}
A particular topic of interest is partial support recovery, and the minimum number of 
samples $M$ necessary for the false positives to vanish in the limit $N\to \infty$. 
Although the fixed point equations \eqref{eq:Q} --\eqref{eq:chih} do not admit a closed form solution, a necessary condition in terms 
of the sample complexity can be derived under the following mild conditions:

\begin{assumption}\label{main_assumption}
  $\quad$
  \vspace{-9pt}
  \begin{enumerate}[label={\textbf{\Alph*:}},
    ref={assumption~\theassumption.\Alph*}]
    \item (Uniqueness of fixed point) The solutions of the fixed point equations \eqref{eq:Q}--\eqref{eq:chih} are unique and satisfy $(Q, \hat{Q}, \chi, \hat{\chi}) \in (0, \infty)^4$.
    \item {\color{black} (Concentration of the oracle Lasso estimator) 
    The random variable \[
      s^{(M)}_{\lambda} := \frac{1}{M} \norm{ \lfit{\lambda}{\bm{y}} - \bm{y} }^2
      \]
    has finite mean $\bar{s}_\lambda^{(M)}$ and variance converging to zero. }
    %\item (Bounded second moment of measurement vector) The measurement vector $\bm{y}$ satisfies $ \norm{\bm{y}} < M$ almost surely.
    \item (Bounded variance of noise distribution) 
    The distribution $p_\xi$ satisfies \[
       \Gamma^{(M)} := \frac{1}{M}\int \dd \bm{\xi} p_{\xi}(\bm{\xi}) \norm{\bm{\xi}}^2 < C
       \] for some constant $C$. 
  \end{enumerate}
\end{assumption}

\begin{claim}[Necessary sample complexity for asymptotically zero false positives]\label{theorem_main}
  
  Let $M$ diverge with $N$ with scaling $M = \alpha \log N  \ (\alpha > 0)$. 
  Under Claim \ref{conj:FE} and Assumption \ref{main_assumption}, 
  if there exists a constant $c > 0$ such that $\av{\FP} < O(N^{-c})$ 
   in the limit $N \to \infty$, then 
   {\color{black}
  \begin{equation}\label{eq:theorem}
      \alpha(1+ \epsilon) > \alpha_C = \frac{\bar{s}_\lambda}{2\lambda^2},
  \end{equation}
  holds for any constant $\epsilon > 0$, where $\displaystyle \bar{s}_\lambda = \lim_{M\to \infty}\bar{s}_\lambda^{(M)}.$}
\end{claim} 
The proof is postponed to Section 4. From this claim, the necessary sample complexity for 
partial support recovery follows immediately:
\begin{claim}[Necessary sample complexity for partial support recovery]\label{theorem_main2}
  Under the settings in Claim \ref{theorem_main},
  if ${\rm supp}(\hat{\bm{x}}_{\lambda}(\vb{A}, \bm{y})) \subseteq {\rm supp}(\bm{x}^0)$ w.a.h.p., then $\alpha > \alpha_C$.  
\end{claim}
{
\color{black}
By definition, $\bar{s}_\lambda$ is the prediction error of the oracle, which 
is given the sensing submatrix $\vb{A}_S$ and observation vector $\bm{y}$. 
This is reminiscent of the primal-dual witness construction in \citet{Wainwright09}, 
where sufficient conditions for asymptotically zero FPs are derived by 
solving the oracle Lasso first, and observing whether the oracle solution concatenated 
with $N-d$ zero elements is a unique solution of the original Lasso problem \eqref{eq:Lasso_estimator}. 
}

Furthermore, the necessary condition for \textit{perfect} support recovery 
can also be derived using Claim \ref{theorem_main}.
\begin{claim}[Necessary sample complexity for perfect support recovery]\label{corollary_perfect}
  Under the settings in Claim \ref{theorem_main}, 
  suppose ${\rm supp}(\hat{\bm{x}}_{\lambda}(\vb{A}, \bm{y})) = {\rm supp}(\bm{x}^0)$ holds w.a.h.p. 
  Then 
  \begin{equation}\label{eq:Wainwright}
   \alpha(1 + \epsilon) > 2\qty(d + \frac{\Gamma}{\lambda^2} ),
  \end{equation}
  holds for any constant $\epsilon > 0$, where $\displaystyle \Gamma = \lim_{M \to \infty} \Gamma^{(M)}$.
\end{claim}
{\color{black}
Note that in the special case of Gaussian noise with variance 
$\sigma^2$, we have $\Gamma = \sigma^2$, 
which extends the result of \citet{Wainwright09}, Theorem 4 
to the case $d = O(1)$. 
Moreover, our result can be applied to any noise distribution satisfying Assumption \ref{main_assumption}.C. 
}
\begin{figure*}[t]
  \begin{centering}
\includegraphics[width = 17cm]{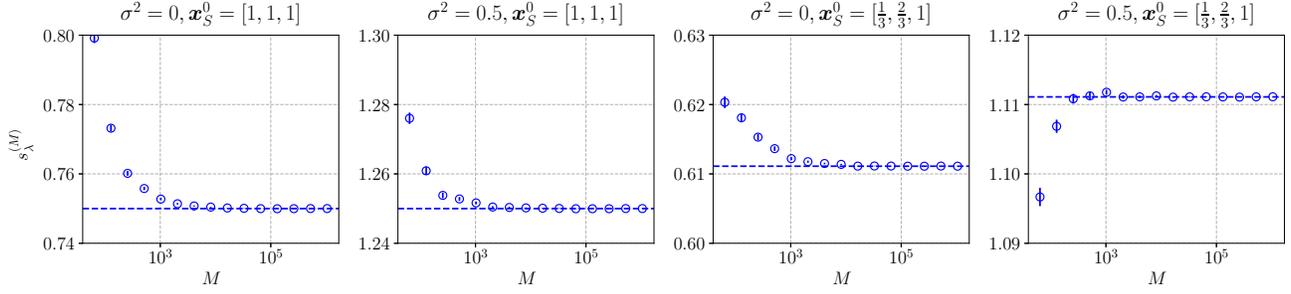}
    \end{centering}
    \caption{ Value of $\bar{s}^{(M)}_\lambda$ for $M$. 
    The asymptotic values of $\bar{s}^{(M)}_\lambda$ (blue dashed lines) are evaluated as $0.750, 1.250, 0.611$ and $1.111$ from left to right. 
    Error bars represent the standard error obtained from 10,000 Monte Carlo samples.
    }
    \label{fig:lasso_finiteM}
  \end{figure*}
  \begin{figure*}[t]
    \begin{centering}
  \includegraphics[width = 17cm]{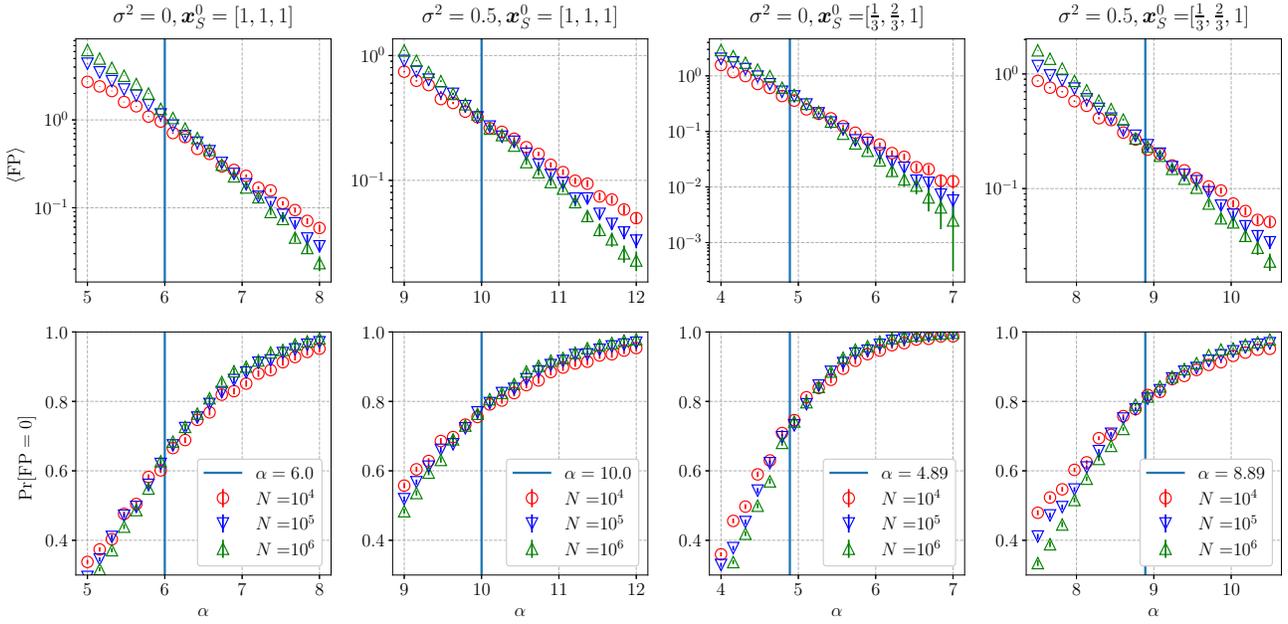}
      \end{centering}
      \caption{Average number of false positives (upper panels) and partial support recovery probability (lower panels)
       near complexity $\alpha = \alpha_C$ (blue vertical lines).
       % for parameters $(\lambda, \sigma^2) = (0.5, 0.0)$ (left) and $(0.5, 0.5)$ (right) at $N = 10^4, 10^5$ and $10^6$. 
      Error bars represent the standard error obtained from 10,000 experimental runs.
      For $\alpha < \alpha_C$, 
      the number of false positives is consistently nondecreasing with respect to $N$, 
      while the partial support recovery probability is consistently nonincreasing with respect to $N$, 
      which is in agreement with Claims \ref{theorem_main} and \ref{theorem_main2}.  }
      \label{fig:lasso_crit}
    \end{figure*}
\section{Numerical experiments}
\subsection{Non-asymptotic results}
To verify the derived results based on Claim \ref{conj:FE}, 
numerical experiments were conducted. For simplicity, we consider the case where the active set 
has size $d = 3$ with $\bm{x}^0_S = \bm{1}_3$, and $\bm{\xi}$ is generated from a Gaussian distribution 
with variance $\sigma^2$. Here, the value of $d$ is taken to be small enough such that finite-size effects 
are nonignorable. 
The values of $\av{\TP}, \av{\FP}$ and $\epsilon_x$
 obtained from our replica predictions 
\eqref{eq:replica_TP}-\eqref{eq:replica_l2} are compared with the average over $10^4$ 
experimental runs. 
The average with respect to $(\bm{A}_S, \bm{y})$ for obtaining the replica prediction 
 was approximated using a Monte Carlo procedure over $10^6$ samples. 

Figure \ref{fig:lasso_general} shows that all three values from theory and experiment are in good agreement for parameters 
$(\lambda, \sigma^2) = (0.5, 0.0)$ and $(0.5, 0.5)$.
\subsection{Asymptotic results}
{\color{black}
Claims \ref{theorem_main} and \ref{theorem_main2} are also verified via numerical experiments; see Supplementary Materials for 
numerical experiments on Claim \ref{corollary_perfect}. 
In order to access the critical point $\alpha_C$ in \eqref{eq:theorem}, Monte Carlo experiments 
 were conducted to evaluate $s_\lambda^{(M)}$ for different values of $M$. 
Figure 2 shows the value of $s_\lambda^{(M)}$ at $(\lambda, \sigma^2) = (0.5, 0.0)$ and $(0.5, 0.5)$ for both 
$\bm{x}_S^0 = \bm{1}_3$ and $\bm{x}_S^0 = [\frac{1}{3}, \frac{2}{3}, 1]$.
 From its asympototic behavior, $\alpha_C$ can be evaluated as the values given in Table 1. 
 Interestingly, for the case $\bm{x}_S^0 = \bm{1}_3$, $s_\lambda$ approaches $6$ and $10$ for 
 $\sigma^2 = 0$ and $0.5$ respectively, which is equivalent to $2(d + \Gamma/\lambda^2)$ given in 
 Claim \ref{corollary_perfect}. 
 
 \begin{table}[h]
  \caption{Values of $\alpha_C$ evaluated from figure 2.} \label{sample-table} 
  \begin{center}
  \begin{tabular}{c|c||c}
  \textbf{$\bm{x}_S^0$}  &\textbf{$(\lambda, \sigma^2)$} & \textbf{$\alpha_C$} \\
  \hline
  $[1,1,1]$        &$(0.5, 0.0)$ &  6.00 \\
  $[1,1,1]$        &$(0.5, 0.5)$ &  10.0 \\
  $[1/3,2/3,1]$    &$(0.5, 0.0)$ &  4.89 \\
  $[1/3,2/3,1]$    &$(0.5, 0.5)$ &  8.89 \\
  \end{tabular}
  \end{center}
  \end{table}

  Figure 3 shows the average number of FP and partial support recovery probability 
  over 10,000 experimental runs for $\alpha$ in the vicinity of the numerically evaluated $\alpha_C$ 
  for different values of $N$. 
  We observe that for $\alpha < \alpha_C$, 
  the average FP is consistently nondecreasing with respect to $N$, while partial support recovery probability 
  is consistently nonincreasing with respect to $N$.  
}
\section{Proofs}

\subsection{Proof of Claim \ref{theorem_main}}
The following lemmas will be useful in the proof. 
\begin{lemma}[Lemma 1, \citet{Dossal12}]\label{lemma_cont}
    There is a finite increasing sequence $(\lambda_t)_{t \leq K}$ with $\lambda_0 = 0$ such that for all $t < K$, 
    the sign and support of $\hat{\bm{x}}_{\lambda}(\vb{A}_S, \bm{y})$ are constant on each interval $(\lambda_t, \lambda_{t + 1})$. 
\end{lemma}
%This is due to Lemma 1 in \citet{Dossal12}. 
\begin{lemma}[Lemma 1, \citet{Tibshirani12}]\label{lemma_Lip}
    The Lasso fit is 1-Lipschitz continuous with respect to $\ell_2$ norm. 
\end{lemma}
%This is due to Lemma 1 in \citet{Tibshirani12}. 
\begin{lemma}[Theorem II.13, \citet{Davidson01}]\label{lemma_eigen}
  Let $\vb{A}\in \mathbb{R}^{M\times d}$ be a random matrix with i.i.d standard Gaussian entries. The 
  largest and smallest eigenvalue of $\vb{B} = \vb{A}^\ten \vb{A}$ satisfy
  \begin{equation}
    \Prob \qty[ \lambda_{\max}(\vb{B}) \geq \qty( \sqrt{M}+ \sqrt{d}+t )^2 ] \leq e^{-\frac{t^2}{2}}
  \end{equation}
   for $t > 0$ and 
  \begin{equation}\label{eq:lam_min}
    \Prob \qty[ \lambda_{\min}(\vb{B}) \leq \qty(\sqrt{M} - \sqrt{d}-t)^2 ] \leq e^{-\frac{t^2}{2}}
  \end{equation}
  for $0 < t < \sqrt{M}-\sqrt{d}$.
\end{lemma}
We now prove Claim \ref{theorem_main}.
Define \[
s^{(M)}_{\lambda, Q} := \frac{1}{M}\int D \bm{z} \norm{\lfit{\lambda}{\bm{y} + \sqrt{Q} \bm{z}} - (\bm{y} + \sqrt{Q}\bm{z}) }^2.
\]
Let us evaluate the difference between 
$s^{(M)}_{(1 + \chi)\lambda, Q}$ and $s^{(M)}_{\lambda, 0} = s^{(M)}_{\lambda}$ when $\av{\FP} < O(N^{-c})$. 
Using the Cauchy Schwartz inequality and symmetry $\lfit{\lambda}{\bm{y}} = -\lfit{\lambda}{-\bm{y}}$,
\begin{equation}
  \label{eq:fitDifference}
\begin{gathered}
        M(s^{(M)}_{(1 + \chi)\lambda, Q} - s^{(M)}_{\lambda}) \\
        \geq -\int D \bm{z} \norm{ \lfit{(1+\chi)\lambda}{\sqrt{Q}\bm{z} + \bm{y}} - \lfit{\lambda}{\bm{y}} - \sqrt{Q}\bm{z} }\\
        \times \norm{ \lfit{(1+\chi)\lambda}{\sqrt{Q}\bm{z} + \bm{y}} - \lfit{\lambda}{-\bm{y}} - 2\bm{y}-\sqrt{Q}\bm{z} }.
    \end{gathered}
  \end{equation}
  The triangle inequality and Lemma \ref{lemma_Lip} implies that 
  \begin{align}\label{eq:first_term}
  &\norm{ \lfit{(1+\chi)\lambda}{\sqrt{Q}\bm{z} + \bm{y}} - \lfit{\lambda}{\bm{y}} - \sqrt{Q}\bm{z} } \nonumber \\
\leq &\sqrt{Q}\norm{\bm{z}} + \norm{ \lfit{(1+\chi)\lambda}{\sqrt{Q}\bm{z} + \bm{y}} - \lfit{(1+\chi)\lambda}{\bm{y}} } \nonumber \\
 & + \norm{ \lfit{(1+\chi)\lambda}{\bm{y}} - \lfit{\lambda}{\bm{y}} }\nonumber \\
 \leq & 2\sqrt{Q}\norm{\bm{z}} + \norm{ \lfit{(1+\chi)\lambda}{\bm{y}} - \lfit{\lambda}{\bm{y}} },
\end{align}
and similarily, 
\begin{align}\label{eq:second_term}
  &\norm{ \lfit{(1+\chi)\lambda}{\sqrt{Q}\bm{z} + \bm{y}} - \lfit{\lambda}{-\bm{y}} - 2\bm{y}-\sqrt{Q}\bm{z} } \nonumber \\
\leq & 2\sqrt{Q}\norm{\bm{z}} + 4\norm{\bm{y}} + \norm{ \lfit{(1+\chi)\lambda}{\bm{y}} - \lfit{\lambda}{\bm{y}} }.
\end{align}
    To derive a bound for the last term in \eqref{eq:first_term} and \eqref{eq:second_term}, Lemma \ref{lemma_cont} is employed.
    Let the support and sign of $\hat{\bm{x}}_{\lambda}(\vb{A}_S, \bm{y})$ be constant in intervals $(M\lambda_t ,M\lambda_{t  + 1}) \ (t = 0, \cdots , K -1)$, 
    where $\lambda = \lambda_0 < \cdots < \lambda_K = (1 + \chi)\lambda$. Let the support set in interval $(\lambda_t , \lambda_{t  + 1})$ 
    be given by $I_t$, and define $\bm{s}_t\in\qty{-1,0,1}^{\abs{I_t}}$ be the sign vector of $\hat{\bm{x}}_{\lambda^\prime}(\vb{A}_S, \bm{y}) \ (\lambda^\prime \in (\lambda_t, \lambda_{t + 1})) $ 
    restricted to $I_t$. From the KKT conditions, the Lasso fit is expressed as 
    \begin{equation*}
        \lfit{\lambda_t}{\bm{y}} = \vb{A}_{SI_t}\vb{A}_{SI_t}^+ \bm{y} - M\lambda_t\vb{A}_{SI_t}(\vb{A}_{SI_t}^\ten \vb{A}_{SI_t} )^{-1}\bm{s}_t, 
    \end{equation*}
    where $\vb{M}^+$ denotes the pseudoinverse of matrix $\vb{M}$. 
    We deduce
    \begin{align*}
        &\norm{ \lfit{(1 + \chi)\lambda}{\bm{y}} -  \lfit{\lambda}{\bm{y}} }  \leq \sum_{t = 0}^{K-1} \norm{ \lfit{\lambda_t}{\bm{y}} -  \lfit{\lambda_{t + 1}}{\bm{y}} } \\      
         \leq &M\lambda \sum_{t = 0}^{K-1} (\lambda_{t + 1} - \lambda_t) \norm{ \vb{A}_{SI_t}(\vb{A}_{SI_t}^\ten \vb{A}_{SI_t} )^{-1}\bm{s}_t } \\ 
        \leq&\chi M \sqrt{d\lambda^2} \max_{t}  \sqrt{\rho((\vb{A}_{SI_t}^\ten \vb{A}_{SI_t} )^{-1})} .
    \end{align*}
    Lemma \ref{lemma_eigen}, with the inclusion principle $\rho((\vb{A}_{SI_t}^\ten \vb{A}_{SI_t} )^{-1}) \leq \rho((\vb{A}_S^\ten \vb{A}_S )^{-1})$ implies that
    w.a.h.p., $\norm{ \lfit{(1 + \chi)\lambda}{\bm{y}} -  \lfit{\lambda}{\bm{y}} }  \leq 2 \chi  \sqrt{d\lambda^2M}$. 
    The relations \eqref{eq:fitDifference} -- \eqref{eq:second_term}, 
    and inequality $\int D \bm{z} \norm{\bm{z}} = \sqrt{2}\Gamma((M+1)/2)/\Gamma(M/2) < \sqrt{M}$ then leads to 
    the following holding w.a.h.p.
    \begin{align}\label{eq:final_inequality}
      &s^{(M)}_{(1+\chi)\lambda, Q} - s^{(M)}_{\lambda} \nonumber \\
      \geq& -4 (\sqrt{Q} + \chi \sqrt{d\lambda^2})\qty(\sqrt{Q} +  \chi \sqrt{d\lambda^2} + \frac{2\norm{\bm{y}}}{\sqrt{M}}).
    \end{align}
    We now use the following lemma which shows that $Q$ and $\chi$ are negligible almost surely.
    \begin{lemma}\label{lemma_smallparams}
      Under the assumptions of Claim \ref{theorem_main}, $\chi < N^{-c/2}$ and $Q < N^{-c/4}$ 
      holds w.a.h.p. 
    \end{lemma}
    The proof is given in Supplementary Materials. 
    Since $\norm{\bm{y}} < \norm{\vb{A}_S \bm{x}^0} + \norm{\bm{\xi}}$ 
    is bounded by $M^2$ w.p.a.1 from 
    Lemma \ref{lemma_eigen} and Assumption \ref{main_assumption}.C, 
    the right hand side of 
    eq. \eqref{eq:final_inequality} is of $O(N^{-c/8 })$ w.p.a.1.
    \color{black}
    We therefore have 
    \begin{equation}\label{eq:two_random_var}
      \Probability\qty[ \frac{\hat{\chi}}{M} = \frac{s^{(M)}_{(1 + \chi)\lambda, Q}}{(1 + \chi)^2} \geq s_\lambda^{(M)} - O(N^{-c/8}) ] > 1 - o(1).
    \end{equation}

    On the other hand, the extremum conditions \eqref{eq:chi} and \eqref{eq:Qhat} imply that 
    $\hat{\chi}$ is always bounded. 
    \begin{lemma}\label{lemma_bounded_chi}
      Suppose the extremum conditions \eqref{eq:Q}--\eqref{eq:chih} are satisfied. Then, 
      the variable $\hat{\chi}$ satsfies 
      \begin{equation}
        \frac{\hat{\chi}}{M} \leq \frac{1}{2}\alpha \lambda^2 \qty(1 -(2\alpha + 1) \frac{\log M}{M} )^{-1}.
      \end{equation}
      
    \end{lemma}
Combined with \eqref{eq:two_random_var}, for sufficiently large $M$
\begin{equation}\label{eq:s_lambda_bound}
  \Probability\qty[ s_\lambda^{(M)} \leq \frac{1}{2}\alpha \lambda^2\qty(1 + \epsilon ) ] > 1 - o(1),
\end{equation}
holds for arbitrary constant $\epsilon > 0$. This implies that
$\frac{1}{2}\alpha \lambda^2(1 + \epsilon)$ must be larger than the median of $s_\lambda^{(M)}$. 
Now, the difference between the median and average is no larger than 
one standard deviation, which is negligible from Assumption 1.B.
This yields the statement of the claim in the limit $M\to \infty$.
\color{black}
  %\subsection{Proof of Lemma \ref{lemma_smallparams} }

\subsection{Proof of Claim \ref{theorem_main2}}
  From Theorem 6 in \citet{Osborne00}, the number of false positives is bounded by $\min (M, N)$. Hence, 
  we have $\av{\FP} < M\times \Prob\qty[\FP \neq 0] = O(N^{-c})$ for some $c > 0$.
   The statement of Claim \ref{theorem_main2} then follows from Claim \ref{theorem_main}.

\subsection{Proof of Claim \ref{corollary_perfect}}

{\color{black} From Claim \ref{theorem_main} and \ref{theorem_main2}, 
it suffices to show that 
\begin{equation}\label{eq:claim5_obj}
  \E s_\lambda^{(M)} > d\lambda^2 + \Gamma_M - o(1),
\end{equation}
}
The KKT conditions imply that w.p.a.1, 
$
  \vb{A}_S \hat{\bm{x}} = \vb{A}_S \vb{A}_S^+ (\bm{y} - M\lambda (\vb{A}_S^+)^\ten \bm{s} ) ,
$
where we abbreviated $\hat{\bm{x}} := \hat{\bm{x}}_\lambda (\vb{A}_S, \bm{y}),$ and $\bm{s} = \sgn{\hat{\bm{x}}}$. 
Therefore, $ \bm{y} - \vb{A}_S \hat{\bm{x}}$ can be decomposed into a sum of two linearly independent vectors 
\begin{equation}\label{eq:residue_decomposition}
  \bm{y} - \vb{A}_S \hat{\bm{x}} = \bm{v} + \bm{v}_\perp,
\end{equation}
where $
  \bm{v} := M\lambda \vb{A}_S(\vb{A}_S^\ten \vb{A}_S)^{-1} \bm{s}, $ 
 $ \bm{v}_\perp := \mathcal{P}_{\ker(\vb{A}_S)} (\bm{y}) = \mathcal{P}_{\ker(\vb{A}_S)} (\bm{\xi})$, and 
 $\mathcal{P}_{\ker(\vb{A}_S)}$ is the projection onto the kernel of $\vb{A}_S$. 
The average of the squared norm of $\bm{v}$ can be evaluated as 
\begin{align}\label{eq:v_vector_bound}
 \E_{\rm eff} \norm{\bm{v}}^2 \geq \E_{\rm eff} \frac{\Lambda d }{\lambda_{\min}( \bm{A}_S^\ten \bm{A}_S )} \geq \frac{M\lambda^2}{ (1 + \sqrt{d/M})^2 },
\end{align}
where the last inequality follows from Jensen's inequality and $\E_{\rm eff} \lambda_{\min}( \bm{A}_S^\ten \bm{A}_S ) \geq (\sqrt{M}-\sqrt{d})^2$ \citep{Davidson01}. 
{\color{black}

To obtain a lower bound on the squared norm of $\bm{v}_\perp$, 
fix the vector $\bm{\xi}$. Noticing that entries of $\vb{A}_S^\ten \bm{\xi} / \norm{\bm{\xi}}^2$ are i.i.d. standard Gaussian,   
the tail bound for $\chi^2$--random variables \citep{Laurent00} implies that for some constant $C > 0$, 
\begin{equation}
  \Probability\qty[ \norm{\vb{A}_S^\ten \bm{\xi}}^2 \leq C \norm{\bm{\xi}}^2 \log M ] \geq 1 - \frac{1}{M}.
\end{equation}
Using this inequality, \eqref{eq:lam_min} with $t = \sqrt{2\log M}$ and the union bound, we have that 
\begin{align}\label{eq:v_perp}
  \E_{\rm eff} &\norm{\bm{v}_\perp}^2  \geq  \E_{\rm eff }\norm{\bm{\xi}}^2 - \E_{\rm eff} \frac{\norm{\vb{A}_S^\ten \bm{\xi}}^2}{\lambda_{\min}(\vb{A}_S^\ten \vb{A}_S)} \nonumber \\
  \geq & \qty(1 - \frac{2}{M})\qty( 1 - \frac{  C\log M }{ (\sqrt{M} - O(\sqrt{\log M }))^2 }) \E_{\rm \bm{\xi}}  \nonumber \norm{ \bm{\xi}}^2 \\
  = &M\Gamma_M (1 - o(1)).
\end{align}
Equation \eqref{eq:claim5_obj} immediately follows from \eqref{eq:v_vector_bound} and \eqref{eq:v_perp}, which completes the proof. 
}

\section{Conclusion}
In this paper, we provided an analysis based on an enhanced replica method for assessing the 
average performance of the Lasso estimator under ultra-sparse conditions. 
Besides, we deduced conditions necessary for support recovery which are derived from 
the oracle Lasso estimator. 
Numerical experiments strongly support the validity of our analysis. 

The methodological novelty originates from 
an observation of finite-size effects and correlations within the active set, 
which is implicitly assumed to be negligible in the conventional replica analysis. 
We anticipate that this framework is applicable to analysis of 
other machine learning or optimization problems where finite-size effects are 
nonnegligible. 
Extending this method further to more general sensing matrix ensembles 
is also another exciting direction for future work. 

\subsubsection*{Acknowledgements}
This work was partially supported by JSPS KAKENHI Grant Nos. 22J21581 (KO), 
21K21310 (TT), 17H00764, 19H01812, 22H05117 (YK) and JST CREST Grant Number 
JPMJCR1912 (YK). 

\subsubsection*{References}

\bibliography{ref}

%%%%%%%%%%%%%%%%%%%%%%%%%%%%%%%%%%%
%%%%%% SUPPLEMENT (OPTIONAL) %%%%%%
%%%%%%%%%%%%%%%%%%%%%%%%%%%%%%%%%%%

\clearpage
\appendix
%\section{FORMATTING INSTRUCTIONS FOR THE SUPPLEMENTARY MATERIAL}
\thispagestyle{empty}
\onecolumn
\aistatstitle{Supplementary Materials}
\newcommand{\eqdef}{\overset{\mathrm{RS}}{=\joinrel=}}
\section{Detailed derivation of Claim \ref{conj:FE}}
Here, we derive the expression in Claim \ref{conj:FE}; see Figure 4 for an outline of the calcuation. 
For simplicity, we abbreviate $\E_{\vb{A}_{\backslash S}}$, the average 
over $\vb{A}$ excluding the submatrix acting on $S$, as $\E$, and $\vb{A}_{\backslash S}$ as the submatrix of $\vb{A}$ excluding $\vb{A}_S$.  
Using the shorthand expression $\dd \bm{x}^a_S := \prod_{i\in S}\dd x_i^a$ and $\dd \bm{\Delta}^a := \prod_{i\neq S} \dd \Delta_i^a$, 
$\E Z^n_\beta(\vb{A}, \bm{y})$ can be written as 
\begin{align}\label{eq:av}
  \E Z^n_\beta(\vb{A}, \bm{y}) &= \int \qty(\prod_{a = 1}^n \dd \bm{x}_S^a \dd \bm{\Delta}^a e^{-\beta M \lambda \norm{\bm{\Delta^a}_1}}) \E \qty[\prod_{a = 1}^n\int \dd \bm{h}^a \delta\qty(\bm{h}^a - \vb{A}_{\backslash S} \bm{\Delta}^a ) e^{-\frac{\beta}{2} \norm{ \bm{A}_S\bm{x}^a_S - \bm{h}^a - \bm{y} }^2 - \beta M \lambda \norm{\bm{x}^a_S}_1} ] 
\end{align}
Using the Fourier representation, the average of the delta functions over $\vb{A}_{\backslash S} = (\tba_1, \cdots, \tba_M)^\ten$ is given by
\begin{gather*}
  \E  \prod_{a = 1}^n \delta \qty( \bm{h}^a - \vb{A}_{\backslash S} \bm{\Delta}^a ) = \int \prod_{\mu = 1}^M  \frac{\dd \tba_\mu e^{-\frac{1}{2} \norm{\tba_\mu}^2}}{(2\pi)^{(N-d)/2}} \prod_{a=1}^n \int  \frac{\dd \tilde{h}_\mu^a}{2\pi} e^{-\iu\tilde{h}_\mu^a(h_\mu^a - \tba_\mu^\ten \bm{\Delta}^a)}\\
  =  \int \prod_{\mu = 1}^M \qty{ \frac{\dd \tba_\mu }{(2\pi)^{(N-d)/2}}  \int \qty( \prod_{a=1}^n \frac{\dd \tilde{h}_\mu^a}{2\pi}) \exp \qty [- \iu\sum_{a = 1}^n h_\mu^a \tilde{h}_\mu ^a-\frac{1}{2} \norm{\tba_\mu - \iu \sum_{a = 1}^n\tilde{h}_\mu^a \bm{\Delta}^a}^2 - \frac{1}{2} \sum_{a,b}\tilde{h}_\mu^a \tilde{h}_\mu^b (\bm{\Delta}^a)^\ten \bm{\Delta}^b ]}\\
 = \int \qty(\prod_{a = 1}^n \prod_{\mu =1}^M \frac{\dd \tilde{h}_\mu^a}{2\pi}) \exp \qty[  - \frac{1}{2}\sum_{\mu = 1}^M \sum_{a,b}\tilde{h}_\mu^a \tilde{h}_\mu^b (\bm{\Delta}^a)^\ten \bm{\Delta}^b  - \iu\sum_{a = 1}^n\sum_{\mu = 1}^M h_\mu^a \tilde{h}_\mu ^a ] =\prod_{\mu = 1}^M \frac{1}{\sqrt{ (2\pi)^n \det \bm{Q} }} \exp \qty( -\frac{1}{2} \bm{h}_\mu^\ten \vb{Q}^{-1} \bm{h}_\mu ),
\end{gather*}
where we defined the matrix $\vb{Q}$ as $\qty(\vb{Q})_{ab} := (\bm{\Delta}^a)^\ten \bm{\Delta}^b$ and used the notation $\bm{h}_\mu := (h_\mu^1, \cdots, h_{\mu}^n)\in \mathbb{R}^n$ 
without confusion. This implies that the vector $\bm{h}_\mu$ is Gaussian with covariance matrix $\vb{Q}$. 
Now, the replica symmetric ansatz \eqref{eq:RS} implies that the integral over $\qty{\bm{\Delta}^a}_{a = 1}^n$ is 
dominated by the subspace of the form 
\begin{equation}\label{eq:subshell_SM}
\int \dd Q \dd \chi \prod_{a = 1}^n \delta\qty(Q - \norm{\bm{\Delta}^a}^2) \prod_{a < b} \delta \qty( Q - \frac{\chi}{\beta} - (\bm{\Delta}^a)^\ten \bm{\Delta}^b ),
%=\int \dd Q \dd \chi \int_{-\iu \infty}^{+\iu \infty}\dd \hat{Q} \dd \hat{\chi} \exp \qty( -\frac{n\beta Q\hat{Q} + n(n-1)\chi \hat{\chi} }{2} - \frac{\beta \hat{Q}}{2}\sum_{a = 1}^n \norm{\bm{\Delta}^a}^2 + \beta^2 \hat{\chi} \norm{\sum_{a= 1}^n \bm{\Delta}^a}^2 + o(\beta) )
\end{equation}
which allows us to simplify the profile of $h_\mu^a$ as 
\begin{equation} \label{eq:local_field}
h_\mu^a = \sqrt{Q - \frac{\chi}{\beta}}z_\mu + \sqrt{\frac{\chi}{\beta}} v^a_\mu,
\end{equation}
where $z_\mu$ and $v_\mu^a \ (a = 1, \cdots, n)$ are all i.i.d. standard Gaussian variables. 
Using \eqref{eq:av}--\eqref{eq:local_field} yields the expression 
\begin{equation}\label{eq:expression_IL}
  \E Z^n_\beta(\vb{A}, \bm{y}) = \int \prod_{a = 1}^n \dd \bm{\Delta}^a \int \dd Q \dd \chi e^{-\beta M \lambda\sum_{a = 1}^n \norm{\bm{\Delta}^a_1}} \mathcal{I}\mathcal{L},
\end{equation}
where $\mathcal{L}$ is given by 
\begin{align}
  \log \mathcal{L} &= \log \int D \bm{z} \prod_{a = 1}^n \qty{ \int D \bm{v}^a \dd \bm{x}^a_S \exp \qty( -\frac{\beta}{2} \norm{\vb{A}_S \bm{x}_S^a - \sqrt{Q - \frac{\chi}{\beta}} \bm{z} - \sqrt{\frac{\chi}{\beta}}\bm{v}^a - \bm{y} }^2 - \beta M \lambda \norm{\bm{x}_S^a}_1 )} \nonumber \\
  & = \log \int D \bm{z} \qty{ \int \dd \vb{x}_S \exp \qty( -\frac{\beta}{2(1 + \chi)} \norm{\vb{A}_S \bm{x}_S - \sqrt{Q}\bm{z} - \bm{y} }^2 - \beta M \lambda \norm{\bm{x}_S}_1 + o(\beta) ) }^n \nonumber \\
  & = n\int D \bm{z} \log \int \dd \vb{x}_S  \exp \qty( -\frac{\beta}{2(1 + \chi)} \norm{\vb{A}_S \bm{x}_S - \sqrt{Q}\bm{z} - \bm{y} }^2 - \beta M \lambda \norm{\bm{x}_S}_1 ) + O(n^2) + o(\beta). \nonumber
\end{align} 

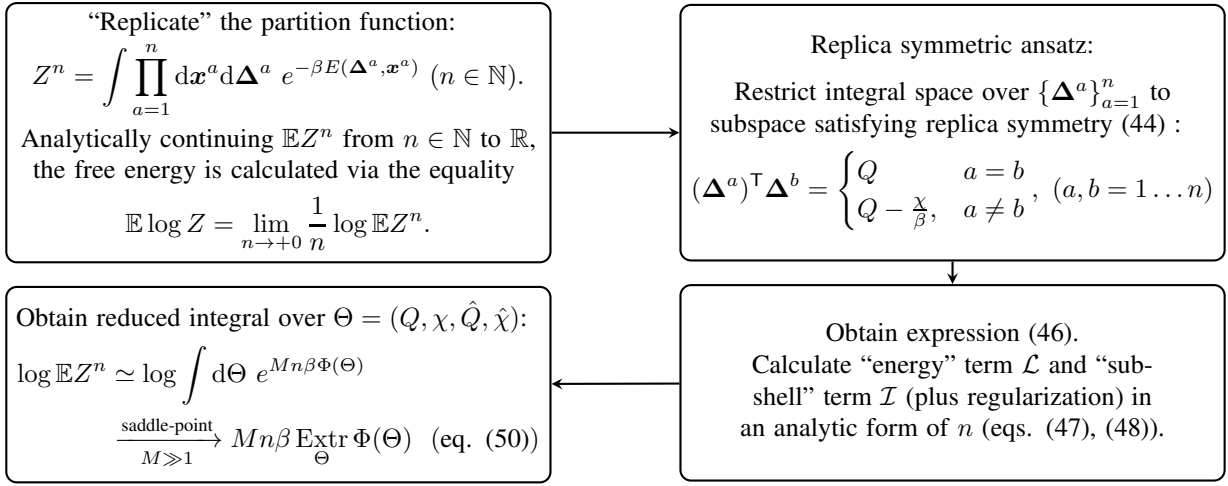
\begin{figure}[t]
  \begin{tikzpicture}
    \node[bblock] (a) {
      ``Replicate" the partition function:
    \vspace{-9pt}
    \begin{equation*}
      Z^n = \int \prod_{a = 1}^n \dd \bm{x}^a \dd \bm{\Delta}^a\ e^{-\beta E(\bm{\Delta}^a, \bm{x}^a)} \ (n \in \mathbb{N}). \vspace{-7pt}
    \end{equation*}
    Analytically continuing $\E Z^n$ from $n \in \mathbb{N}$ to $\mathbb{R}$, the free energy 
    is calculated via the equality
    \vspace{-6pt}
    \begin{align*}
      \E \log Z &=  \lim_{n\to +0} \frac{1}{n}\log \E Z^n.
    \end{align*}
    };
    \node[bblock, right=of a, xshift=0.7cm] (d) {
      Replica symmetric ansatz:\\
      \vspace{6pt}
      Restrict integral space over $\qty{\bm{\Delta}^a}_{a  =1}^n$ to subspace satisfying replica symmetry \eqref{eq:subshell_SM} :
      \vspace{-6pt}
      \begin{equation*}
        (\bm{\Delta}^a)^\ten \bm{\Delta}^b = 
        \begin{cases}
          Q & a = b\\
          Q - \frac{\chi}{\beta}, & a \neq b
        \end{cases}, \ (a, b = 1 \myDots n)
      \end{equation*}
    };
    \node[block, below=of d, yshift=0.7cm] (c) { 
      Obtain expression \eqref{eq:expression_IL}. \\
      Calculate ``energy" term $\mathcal{L}$ and ``subshell" term $\mathcal{I}$ (plus regularization) 
      in an analytic form of $n$ (eqs. \eqref{eq:expression_L}, \eqref{eq:expression_I}).
  };
  \node[block, below=of a, yshift=0.7cm] (b) {
      Obtain reduced integral over $\Theta = (Q, \chi , \hat{Q}, \hat{\chi})$:
      \vspace{-6pt}
      \begin{align*}
       \log \E Z^n &\simeq \log \int \dd \Theta \ e^{Mn\beta \Phi(\Theta)} \\
       &\xrightarrow[M \gg 1]{\text{saddle-point}}  Mn\beta \Extr_{\Theta} \Phi(\Theta)\ \ (\text{eq. \eqref{eq:Final_expression}})
       \end{align*}
  };
    \draw [arrow] (a) -- (d);
    \draw [arrow] (d) -- (c);
    \draw [arrow] (c) -- (b);
  \end{tikzpicture}
  \caption{Outline of the replica calculation for Claim 1.}
\end{figure}
The integral with respect to $\bm{x}_S$ can be evaluated using Laplace's method for large $\beta$, yielding
\begin{equation}\label{eq:expression_L}
 \mathcal{L} \simeq \exp \qty[ -Mn \beta \int D \bm{z} \min_{\bm{x}} \qty( \frac{\norm{ \vb{A}_S\bm{x} - \sqrt{Q} \bm{z} - \bm{y} }^2}{2M(1+\chi)} + \lambda \norm{\bm{x}}_1 )],
\end{equation}
where the subleading terms are ignored.
Similarily, $\mathcal{I}$ is given by 
\begin{align}
  \mathcal{I} &= \prod_{a = 1}^n \delta\qty(Q - \norm{\bm{\Delta}^a}^2) \prod_{a < b} \delta \qty( Q - \frac{\chi}{\beta} - (\bm{\Delta}^a)^\ten \bm{\Delta}^b ) \nonumber \\
  & = \int_{-\iu \infty}^{+\iu \infty}\dd \hat{Q} \dd \hat{\chi} \exp M \qty[ \frac{1}{2}(\beta\hat{Q} - \beta^2\hat{\chi} ) \sum_{a = 1}^n\qty(Q - \norm{\bm{\Delta}^a}^2) - \frac{1}{2}\beta^2\hat{\chi} \sum_{a\neq b}\qty( Q - \frac{\chi}{\beta} - (\bm{\Delta}^a)^\ten \bm{\Delta}^b ) + o(\beta) ] \nonumber\\
  & = \int_{-\iu \infty}^{+\iu \infty}\dd \hat{Q} \dd \hat{\chi} e^{\frac{n\beta M}{2} \qty( Q\hat{Q} + (n-1)\chi \hat{\chi} - n\beta Q\hat{\chi} )} \exp M \qty[ -\frac{\beta\hat{Q}}{2} \sum_{a = 1}^n \norm{\bm{\Delta}^a}^2 + \frac{\beta^2\hat{\chi}}{2} \norm{\sum_{a = 1} \bm{\Delta}^a}^2 + o(\beta) ] \nonumber \\
  & =  \int_{-\iu \infty}^{+\iu \infty}\dd \hat{Q} \dd \hat{\chi} e^{\frac{n\beta M}{2} \qty( Q\hat{Q} + (n-1)\chi \hat{\chi} - n\beta Q\hat{\chi} )} \int D \hat{\bm{z}}  \exp \beta \qty[ \qty(-\frac{ M \hat{Q}  }{2} \sum_{a = 1}^n \norm{\bm{\Delta}^a}^2 + \sqrt{M\hat{\chi}} \sum_{a= 1}^n \hat{\bm{z}}^\ten \bm{\Delta}^a) + o(\beta) ]. \nonumber
\end{align}
Therefore, ignoring the subleading term with respect to $\beta$, 
\begin{align}\label{eq:expression_I}
  & \int \prod_{a = 1}^n \dd \bm{\Delta}^a e^{-\beta M \lambda\sum_{a = 1}^n \norm{\bm{\Delta}^a_1}} \mathcal{I}  \nonumber \\
  \simeq & \int_{-\iu \infty}^{+\iu \infty}\dd \hat{Q} \dd \hat{\chi} e^{\frac{n\beta M}{2} \qty( Q\hat{Q} + (n-1)\chi \hat{\chi} - n\beta Q\hat{\chi} )}\int D \hat{\bm{z}} \qty{\int \dd \bm{\Delta} \exp \beta \qty[ -\frac{M \hat{Q}}{2} \norm{\bm{\Delta}}^2 + \sqrt{M\hat{\chi}} \hat{\bm{z}}^\ten \bm{\Delta} - M \lambda \norm{\bm{\Delta}}_1  ]}^n
\end{align}
The log of the integral with respect to $D\hat{\bm{z}}$ can be expanded as 
\begin{align}
  &\log \int D\hat{\bm{z}} \qty{\int \dd \bm{\Delta} \exp \beta \qty[ -\frac{M \hat{Q}}{2} \norm{\bm{\Delta}}^2 + \sqrt{M\hat{\chi}} \hat{\bm{z}}^\ten \bm{\Delta} - M \lambda \norm{\bm{\Delta}}_1  ]}^n \nonumber \\
  =&\ n \sum_{i = 1}^{\tN}  \int D \hat{z}_i \log \int \dd \Delta \exp \beta \qty[ -\frac{M \hat{Q}}{2} \Delta^2 + \sqrt{M\hat{\chi}} \hat{z}_i \Delta - M \lambda \abs{\Delta}  ] + O(n^2) \nonumber \\
  \simeq &\ - n \beta \tN \int D \hat{z} \min_{\Delta} \qty( \frac{M \hat{Q}}{2} \Delta^2 - \sqrt{M\hat{\chi}} \hat{z} \Delta + M \lambda \abs{\Delta}  ) + O(n^2) \nonumber \\
  \label{eq:exp_of_I}=&\ Mn\beta \frac{\tN }{2\hat{Q}} \qty[ \qty( \lambda^2 + \frac{\hat{\chi}}{M}) \erfc\qty(\sqrt{\frac{\Lambda}{2M\hat{\chi}}} ) - \sqrt{ \frac{2\lambda^2 \hat{\chi} }{\pi M}} e^{-\Lambda / 2M\hat{\chi}} ] +O(n^2) ,
\end{align}
where Laplace's approximation was used for large $\beta$ to obtain the third line.
Substituting \eqref{eq:expression_L}, \eqref{eq:expression_I} and \eqref{eq:exp_of_I} 
into \eqref{eq:expression_IL}, using the saddle point method for large $M$ results in 
\begin{equation}\label{eq:Final_expression}
  \begin{gathered}
  \log \E Z^n_\beta(\vb{A}, \bm{y}) = Mn\beta \Extr_{Q, \hat{Q}, \chi, \hat{\chi}} \Bigg\{ \frac{Q\hat{Q} + (n-1)\chi \hat{\chi} - n\beta Q \hat{\chi}}{2} \\
  -\int D \bm{z} \min_{\bm{x}} \qty( \frac{\norm{ \vb{A}_S\bm{x} - \sqrt{Q} \bm{z} - \bm{y} }^2}{2M(1+\chi)} + \lambda \norm{\bm{x}}_1 ) + \frac{\tN }{2\hat{Q}} \qty[ \qty( \lambda^2 + \frac{\hat{\chi}}{M}) \erfc\qty(\sqrt{\frac{\Lambda}{2M\hat{\chi}}} ) - \sqrt{ \frac{2\lambda^2 \hat{\chi} }{\pi M}} e^{-\Lambda / 2M\hat{\chi}} ] \Bigg\}.
  \end{gathered}
\end{equation}
Noticing that 
\begin{equation}
 \lim_{n\to +0}\frac{\E Z_\beta^n(\bm{A}, \bm{y}) -1}{n} =  \lim_{n\to +0}\frac{ \log \E Z_\beta^n(\bm{A}, \bm{y})}{n},
\end{equation}
 and finally rescaling $\hat{Q} \leftarrow M\hat{Q}$ and $\hat{\chi} \leftarrow M\hat{\chi}$, 
one obtains \eqref{eq:FE}.

\section{Proof of auxiliary lemmas}
\subsection{Proof of Lemma 4}

From \eqref{eq:chi} and \eqref{eq:Qhat}, we have 
$
  \chi =$$ f\qty( \frac{\tN}{M-\bar{d}} \erfc \qty( \sqrt{\frac{\Lambda}{2\hat{\chi}}}) ), $
where $f(x) = x/(1-x)$, and $\bar{d}$$ := \int D \bm{z} \norm{ \hat{\bm{x}}_{(1+\chi)\lambda}( \sqrt{Q}\bm{z} + \bm{y}  ) }_0$. 
From $\chi > 0$, we see that $f$ is a increasing function. 
By using the Markov inequality, it can be deduced that 
\[
  \Prob\qty[ \chi > N^{-c_1} ] = \Prob\qty[f^{-1}(\chi) > f^{-1}(N^{-c_1}) ] 
  \leq \frac{\E f^{-1}(\chi)}{f^{-1}(M^{-1})} < \frac{1 + N^{c_1}}{M-d} \av{\FP} < O(N^{-(c - c_1)}), \nonumber
\]
which proves the first part of the lemma with $c_1 = c/2$. For the probability bound on $Q$, using $\erfc(x) < \frac{1}{x\sqrt{\pi}}e^{-x^2}$, 
\begin{align*}
  Q &< \frac{ \hat{\chi}(1 + \chi) }{(M-\bar{d})^2}\tN \erfc \qty( \sqrt{\frac{\Lambda}{2\hat{\chi}}}) = g \qty( \frac{\tN}{M-\bar{d}} \erfc \qty( \sqrt{\frac{\Lambda}{2\hat{\chi}}}) ),
\end{align*}
where 
\begin{align*}
  g(x) &= \frac{M^2\lambda^2}{2(M-d)} \frac{x}{(1-x)^2}\qty[ \erfc^{-1}\qty( \frac{Mx}{N} ) ]^{-2}.
\end{align*}
Using $\erfc^{-1}(Mx/N) > (1-x)^2 $ for $M/N < 1$, we have for large enough $M$,
\begin{equation*}
  g(N^{-c_1}) < \frac{\lambda^2}{2}N^{-c_1} M^{2}(M-d)^{-1}(1-N^{-3c_1})^{-6}.
\end{equation*} 
Since both $g$ and $g^{-1}$ are nonnegative and increasing, for large enough $M$,
\begin{equation*}
  g^{-1}(N^{-c_1/2}) > g^{-1}\qty( \frac{\lambda^2}{2}N^{-c_1} M^{2}(M-d)^{-1}(1-N^{-3c_1})^{-6} ) > N^{-c_1}.
\end{equation*}
The Markov inequality then implies the second part of the lemma with $c_1 = c/2$:
\begin{equation}
\begin{gathered}
  \Prob\qty[ Q > N^{-c_1/2} ] = \Prob\qty[g^{-1}(Q) > g^{-1}(N^{-c_1/2}) ] < \frac{\av{\FP}}{(M-d)g^{-1}(N^{-c_1/2})} < \frac{1}{M-d}N^{-c_1}\av{\FP} < O(N^{-(c - c_1)}) \nonumber.
\end{gathered}
\end{equation}
{\color{black}
\subsection{Proof of Lemma 5}
Equations \eqref{eq:chi} and \eqref{eq:Qhat} imply that $ N\erfc\qty(\frac{\Lambda}{\sqrt{2\hat{\chi}}}) = \hat{Q}\chi \leq M \frac{\chi}{1 + \chi} \leq M$ holds for 
any $(\vb{A}_S, \bm{y})$. Thus, $\hat{\chi}$ is deterministically upper-bounded as 
 \begin{equation}\label{eq:bound_on_chih}
  \hat{\chi} \leq \frac{1}{2}  \frac{\Lambda^2}{ \qty[ \erfc^{-1}(M / N) ]^2 }.  
 \end{equation}
Now, $\erfc$ satisfies \citep{Chang11} for $0 < \epsilon < 1/3$,
\begin{align*}
  \erfc(x) &\geq \exp\qty[ -(1 + \epsilon)x^2 + \log \epsilon ]. \quad  \therefore  \qty[ \erfc^{-1}(x) ]^{-2} \leq (1+\epsilon) ( - \log x + \log\epsilon)^{-1} .
\end{align*}
Applying this inequality to \eqref{eq:bound_on_chih} with $\epsilon = M^{-1}$ and $N = \exp (M / \alpha)$ for $M > 3$ yields 
\begin{equation}
  \hat{\chi} \leq \frac{1}{2} \alpha M\lambda^2\qty(1 + \frac{1}{M})\qty(1-2\alpha \frac{\log M}{M} )^{-1} < \frac{1}{2}\alpha M \lambda^2 \qty( 1 -(2\alpha + 1)\frac{\log M}{M} )^{-1}.
\end{equation}
}

%\section{Additional numerical experiments}
\begin{figure}
  \begin{centering}
  \includegraphics[width = 17cm]{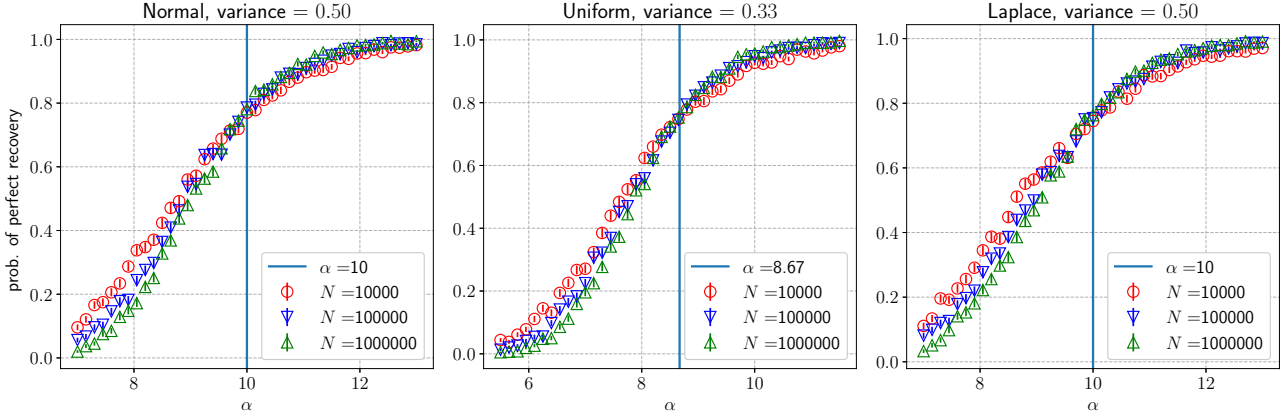}
    \end{centering}
    \caption{Probability of perfect recovery for noise distributed according to (1) Gaussian distribution with mean $0$ and variance $0.5$ (left), 
    (2) Uniform distribution defined on the interval $\qty[-1, 1]$ (middle), and (3) Laplace distribution with mean $0$ and variance $0.5$ (right) at $N = 10^4, 10^5,$ and $10^6$.
    Error bars represent the standard error obtained from 1,000 experimental runs. The horizontal blue line depicts 
    the necessary sample complexity given by Claim \ref{corollary_perfect}. }
    \label{fig:perfect_recovery}
  \end{figure}
  
\section{Additional numerical experiments : Necessary condition for perfect support recovery}
To verify the necessary sample complexity for perfect recovery given by Claim \ref{corollary_perfect}, numerical experiments were 
conducted. The profile of $\bm{x}^0$ is the same as that of Section 3.1, and the regularization parameter is taken as $\lambda = 0.5$. 
Figure \ref{fig:perfect_recovery} shows the perfect support recovery probability for noise distributed according to 
the Gaussian, uniform, and Laplace distribution. Clearly, for all three cases, perfect recovery fails with finite probability as $N$ tends to infinity
 when $\alpha$ is less than the value indicated by Claim \ref{corollary_perfect}. 

\end{document}